# A modular table-top setup for ultrafast X-ray diffraction

W. Lu,[1] M. Nicoul,[1] U. Shymanovich,[1] A. Tarasevitch,[1] M. Horn-von Hoegen[1], D. von der Linde,[1] and K. Sokolowski-Tinten[1,a)]

[1]*Faculty of Physics and Centre for Nanointegration Duisburg-Essen, University of Duisburg-Essen, Lotharstrasse 1,*

*47048 Duisburg, Germany*

We present a table-top setup for femtosecond time-resolved X-ray diffraction based on a Cu $K_\alpha$ (8.05 keV) laser driven plasma X-ray source. Due to its modular design it provides high accessibility to its individual components (e.g. X-ray optics and sample environment). The $K_\alpha$-yield of the source is optimized using a pre-pulse scheme. A magnifying multilayer X-ray mirror with Montel-Helios geometry is used to collect the emitted radiation, resulting in a quasi-collimated flux of more than $10^5$ Cu $K_\alpha$ photons/pulse impinging on the sample under investigation at a repetition rate of 10 Hz. A gas ionization chamber detector is placed right after the X-ray mirror and used for normalization of the diffraction signals enabling the measurement of relative signal changes of less than 1% even at the given low repetition rate. Time-resolved diffraction experiments on laser-excited epitaxial Bi films serve as an example to demonstrate the capabilities of the set-up. The set-up can also be used for Debye-Scherrer type measurements on poly-crystalline samples.

## I. INTRODUCTION

The rapid development of ultrashort pulsed X-ray sources in recent decades has lead to spectacular progress in the field of ultrafast structural dynamics[1,2]. The current state-of-the-art is set by accelerator-based X-ray free electron lasers (XFELs)[3-5]. While the extreme brilliance of these costly, large-scale-facility sources has opened up completely new possibilities, access is restricted and very competitive. Therefore, laboratory-scale, table-top X-ray sources, like short pulse laser-driven plasmas, still represent an interesting alternative due to their low cost (as compared to XFELs), simplicity, versatility and accessibility. In fact, the possibility to generate ultrashort hard X-ray pulses by using high intensity femtosecond laser pulses as a driver was one of the important enabling steps to establish the field. Many important ultrafast diffraction experiments have been successfully carried out with these sources[6-22], and since the first demonstration of sub-ps time resolution[6] they have and are still been further developed, for example towards higher repetition rate[23,24], better efficiency[25,26] or to reach higher photon energies[27,28].

___________________________

a) Author to whom correspondence should be addressed. Electronic mail: klaus.sokolowski-tinten@uni-due.de.

In a simplified view the emission mechanism and the properties of ultrashort X-ray pulses from a fs laser driven plasma created at the surface of a solid target can be compared to X-ray production in an ordinary X-ray tube. However, instead of the static electric field in an X-ray tube, electrons of sufficiently high energy (tens of keV) are generated through direct interaction of *thermal* plasma electrons with the high electrical field of the laser pulse (e.g. via resonance absorption) at intensities of $10^{16} - 10^{18}$ W/cm$^2$. A fraction of those laser accelerated, *hot* electrons propagate into the cold, unexcited material underneath the surface plasma layer generating Bremsstrahlung and characteristic line emission (e.g. $K_\alpha$-emission). Therefore, the emitted X-ray pulses exhibit a short duration, which is determined by the duration of the driving laser pulse and the subsequent stopping time inside the material[29,30]. Moreover, these pulses are perfectly suited for pump-probe type measurements since they are inherently synchronized to the drive laser. For time-resolved diffraction experiments an optical pump-pulse is used for excitation while the ultrashort X-ray pulse monitors the transient structural changes induced by the pump by measuring the diffraction from the excited sample as a function of the time delay between optical pump and X-ray probe.

However, this type of source shares also the spatial properties of an ordinary X-ray tube, namely its incoherent emission into the full solid angle. Therefore, an appropriate X-ray optic, which collects and transfers an as high as possible amount of the emitted radiation onto the sample is almost mandatory[31-37].

In this work, we present a compact and modular table-top setup based on a laser plasma Cu $K_\alpha$ X-ray source for time-resolved X-ray diffraction experiments. It employs a pre-pulse scheme to enhance X-ray production[38], a magnifying Montel multilayer X-ray mirror[39] to deliver a monochromatic and quasi-collimated beam to the sample, and a gas ionization detector for signal normalization. Its application is demonstrated by resolving femtosecond structural dynamics in a laser-excited thin Bismuth film.

## II. MODULAR SETUP FOR TIME-RESOLVED DIFFRACTION

The goal of this work was to construct a flexible and at the same time compact platform for time-resolved X-ray diffraction experiments. As such it employs a design that separates the key components (e.g. X-ray source, X-ray optics, sample environment, and detector) into different modules, which can be individually modified/adapted to particular applications. A scheme of the setup is shown in Fig. 1.

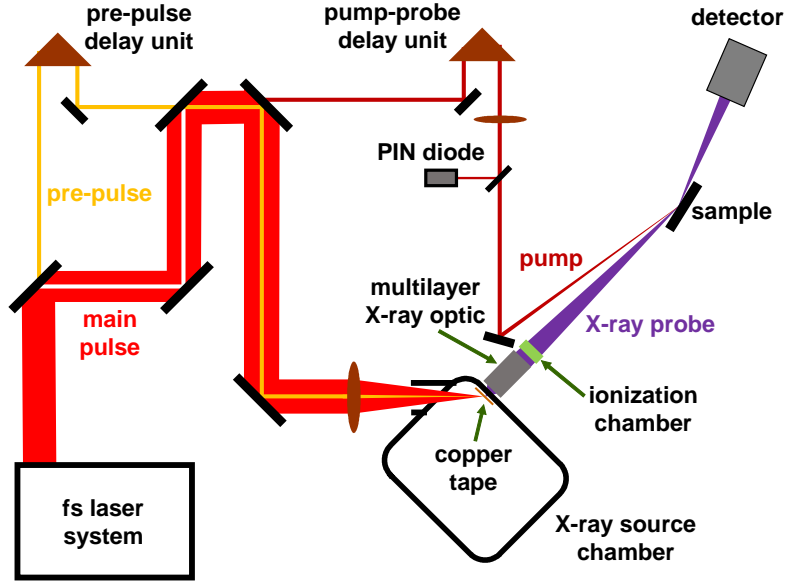

FIG. 1. Scheme of the modular setup based on laser plasma Cu Kα X-ray source for time-resolved X-ray diffraction experiments.

The laser driver for X-ray generation is a chirped-pulse-amplification Titanium Sapphire laser system providing 120 fs pulses (stretched on purpose; see below) at a wavelength of 800 nm with a high laser-pulse contrast ratio (LPCR: $10^7$ at 2 ps ahead of the pulse peak; $>10^8$ with respect to amplified spontaneous emission, ASE). Unlike many of the currently operating laser-driven plasma X-ray sources, which use 5 – 10 mJ, kHz repetition rate laser systems, we have opted for a high energy (150 mJ/pulse), low repetition rate (10 Hz) system. While the average X-ray flux ($\approx 10^6$ $K_\alpha$-photons/s delivered to the sample; details see below) is comparable to these high repetition rate sources, the per-pulse flux is correspondingly (factor 100) higher.

Therefore, the same accumulated X-ray diffraction signal from a given sample can be obtained with less pulses. This enables experiments at relatively high optical pump fluences since optical damage at solid surfaces is often determined by accumulative effects and thus by the total optical dose. In consequence, less number of required pump-probe events automatically allows for higher pump fluences approaching or even exceeding the single-pulse melting threshold[11].

The main laser beam (the red optical path in Fig.1) is split into three parts. A first holey mirror with a small central hole splits off approximately 1.5 mJ of the main pulse energy, which serves as a plasma-generating pre-pulse to enhance X-ray generation[38]. It is send through an optical delay line (pre-pulse delay unit in Fig. 1) and then recombined with the main beam by a second holey mirror. Pre- and main pulse are focused onto the surface of the X-ray target (copper tape) under an angle of incidence of 45° by a 30 cm focal length lens to intensities of $2\times10^{14}$ W/cm$^2$ and $1.6\times10^{17}$ W/cm$^2$, respectively. Due to the smaller near-field diameter of the pre-pulse beam its focus (approx. 80 µm FWHM) is larger than that of main beam (25 µm FWHM) ensuring that the main pulse interacts with a laterally homogeneous pre-plasma.

After the recombining second holey mirror a third holey mirror with an off-center hole splits off another fraction (about 1 mJ) from the main beam, which serves as the optical pump beam to excite the sample. It is focused with a lens of 1 m focal length onto the sample surface. The spot size of the focused pump beam can be adjusted by a diaphragm and monitored by a CCD camera positioned at a reference position equivalent to the sample plane. The energy of the pump beam is adjusted with a half-wave-plate in conjunction with a thin film polarizer and monitored with a PIN diode. This configuration can provide excitation fluences of more than 100 mJ/cm$^2$, which at 800 nm is close to or beyond the melting threshold of many (in particular metallic) materials. A second optical delay line allows to vary the time delay between the optical pump and the X-ray probe beam. The angle between laser pump and X-ray probe beam is set to be as small as possible (<10°).

The X-ray source is the only module that needs to operate under vacuum to avoid degradation of the spatial beam profile of the focused laser beam due to nonlinear effects in air. Therefore, the X-ray target (Cu tape) is mounted inside a vacuum chamber (working pressure 0.1 mbar). Part of the radiation emitted by the laser-generated plasma into the full solid angle leaves the vacuum chamber through a small hole covered with a thin Kapton foil. A multilayer X-ray optic is used to collect the radiation and to refocus it onto the sample under investigation.

A small ionization chamber detector (ICD) is located directly after the X-ray multilayer mirror. It is filled with Ar at a pressure of about 1 bar and the charge signals generated by X-ray induced ionization are used as a reference to normalize the diffraction signals on a pulse-to-pulse basis.

The sample under study needs to be placed in the focus of the multilayer X-ray optic. Depending on the particular application samples can be mounted on dedicated motorized sample stages which different linear and rotational degrees of freedom. This allows to adjust the sample orientation with respect to the incoming X-ray beam as well as the beam position on the sample. Point as well as area X-ray detectors are available to record the X-ray signal diffracted/scattered by the sample.

In order to minimize losses by absorption of the Cu K$_\alpha$-radiation in air, the optics housing is purged with He, and He-purged beam-tubes (not shown in Fig. 1) are installed in the X-ray beam path towards the sample and the detector.

## III. CHARACTERIZATION AND PERFORMANCE

In this section we provide a detailed characterization of the different modules (source, X-ray optic, ICD, target environment, detectors) with particular emphasis on the relevant parameters that define the overall capabilities of the setup. We demonstrate that a full start-to-end design is key to achieve high flexibility and performance.

### A. X-ray source

The vacuum chamber housing the X-ray source is very compact (size 24×30×30 cm$^3$) and made from stainless steel, as shown in Fig. 2(a). To enhance mechanical stability and to provide radiation shielding already close to the source, the chamber exhibits a wall thickness of 15 mm. 2 mm of lead are added to the inside of the chamber walls to further enhance radiation shielding and to reduce also the level of background radiation that may reach the detector. Except in the direction of the X-ray probe beam, which exits the chamber through a 2 mm diameter Kapton window, this ensures a radiation level of less than 1 µSv/h at 10 cm distance from the chamber outer walls.

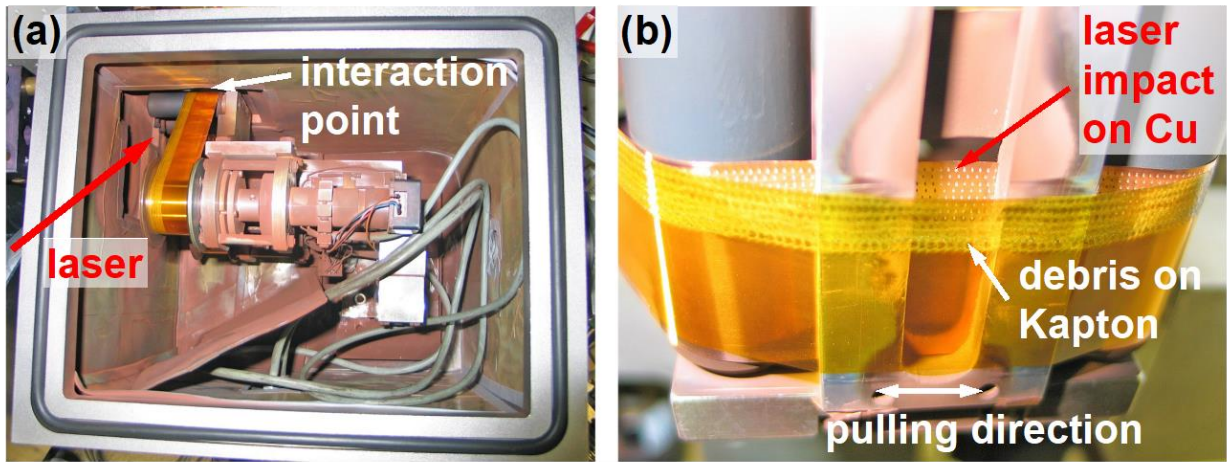

FIG. 2. X-ray source vacuum chamber. (a) Inside view with the installed tape target assembly. Note the large amounts of Cu debris caused by ablation from Cu-tape. (b) Interaction area on the Cu tape target with debris catcher foil (Kapton).

The target consists of a 10 µm thick Cu tape mounted on a motorized spooling system, which has been developed at the Institute for Quantum Electronics of the Friedrich-Schiller-University Jena[40]. The tape is continuously pulled with a speed of approximately 8 mm/s to provide a non-irradiated surface area on the Cu tape for each impinging laser pulse. When the tape reaches its end the pulling direction changes automatically and the tape is shifted at the same time perpendicular to its pulling direction by about 0.8 mm. A typical loading (Cu tape with 25 mm width and 15 m length) can provide more than 12 hours of measurement time.

Debris from ablated material represents a major challenge and requires regular cleaning of the chamber interior and the spooling system. Ablation occurs predominantly by spallation of material at the back surface of the Cu tape upon reflection of the laser-induced shock wave[41]. To directly catch most of the ejected material, a 8 μm thick Kapton foil is mounted together with the Cu tape to the spooling system, so that they are simultaneously pulled with the same velocity. Close to the laser-target interaction point the catcher foil is mechanically separated from the Cu tape by about 2 mm (see Fig.2(b)).

To spectrally characterize the X-ray emission from the source a thinned, back-illuminated CCD detector (Princeton Instruments MTE: 1300B) was used in photon counting mode, so that a pulse-height analysis of the detected single-photon events provides the spectrum of the detected radiation[42,43]. Fig. 3(a) shows a typical single-pulse spectrum obtained in this way.

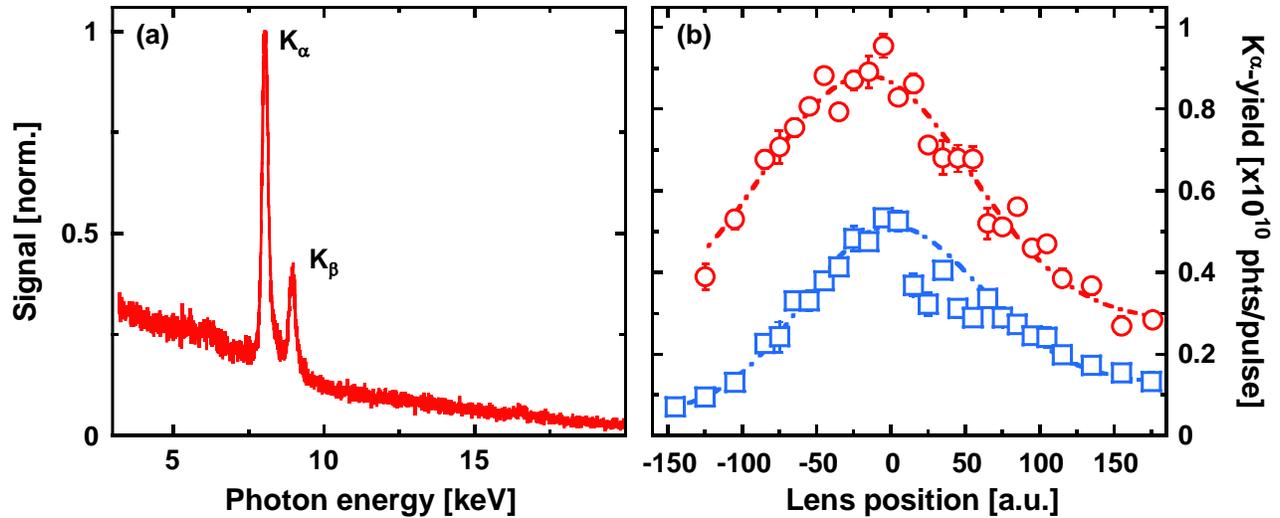

FIG. 3. (a) Single-pulse emission spectrum of the X-ray source. (b) Measured $K_\alpha$-yield (norm.) as a function of the position of the focusing lens (relative to focus position) without (blue squares) and with a pre-pulse (arriving 5 ps before the main pulse; red circles); the dashed-dotted curves represent guides to the eye.

The spectrum exhibits characteristic line radiation (e.g. $K_\alpha$ and $K_\beta$) on a continuous Bremsstrahlung background. Please note that part of the signal at lower energies (below the K-lines) is actually caused by $K_\alpha$ and $K_\beta$ photons with the X-ray generated charge distributed over more than one camera pixel[43]. For the diffraction experiments we employ the $K_\alpha$-emission at 8.05 keV, which consists of the spin-orbit split $K_{\alpha 1}$ and $K_{\alpha 2}$ doublet (not resolved in Fig. 3).

The $K_\alpha$-yield depends on laser (intensity) as well as plasma (scale length of the spatial density profile) parameters[30,38]. Therefore, to optimize $K_\alpha$-generation we vary the laser intensity by adjusting the laser pulse duration and the position of the focusing lens relative to the X-ray target. The plasma scale length is controlled through a suitable pre-pulse. Due to the high

LPCR we are in this way able to separate the processes of plasma formation/expansion and X-ray generation in a highly controllable fashion.

Fig. 3(b) shows the measured $K_\alpha$-yield (norm.) as a function of the relative lens position without pre-pulse (blue squares) and with a pre-pulse (red circles) arriving 5 ps before the main pulse. It can be clearly seen that proper adjustment of the lens position allows to maximize the $K_\alpha$-yield, a procedure that is regularly performed to maintain the source performance. In addition, to allow at the same time for a small X-ray source size the optimum working point should be close to the actual focus position. Therefore, the laser pulses are on purpose not compressed to their bandwidth limit (<50 fs), but stretched pulses of about 100 – 120 fs duration are used.

The data shown in Fig. 3(b) also clearly show the effect of the pre-pulse, which allows to increase the $K_\alpha$-yield by about a factor of 2. Under optimized conditions and assuming spatially isotropic X-ray emission we achieve a $K_\alpha$-flux of $4\times10^9$ photons/s/sr without pre-pulse, and $8\times10^9$ photons/s/sr with pre-pulse at 110 mJ of energy in the main pulse impinging on the Cu tape (corresponding to 1.1 W of average drive laser power on target). This compares well will the reported yield of higher repetition rate sources and a few W drive power at 800 nm[24,26,44].

**B. Multilayer X-ray optic**

As already mentioned, the X-ray emission from laser-produced plasmas occurs incoherently into the full solid angle. Therefore, a suitable X-ray optic is required to collect as much radiation as possible and guide it to the sample. For our setup we have chosen a magnifying multilayer optic with Montel geometry[39] (manufactured by Rigaku-OSMIC Inc.). With a working distance of 100 mm to the source it can be installed outside the vacuum chamber. The mirror exhibits a capture angle of 1.51° and a magnification of 5, resulting in a mirror-to-sample distance of 500 mm. The large mirror-to-sample distance gives sufficient space and thus great flexibility with respect to the sample environment (see section D). The multilayer coating is designed to reflect both the Cu $K_{\alpha1}$ and $K_{\alpha2}$ emission lines, but suppresses other spectral components (i.e. $K_\beta$). In order to minimize losses by absorption of the Cu $K_\alpha$-radiation in air, the housing of the Montel-optic is purged with He.

Fig. 4 presents results on the spatial characterization of the mirror. The image shown in Fig. 4(a) has been recorded with the X-ray CCD placed after the X-ray mirror (with the ICD removed) at a distance of 40 cm from the focus, and the X-ray flux attenuated by suitable Al-filters. The beam exhibits a quadratic shape of 1.6 x 1.6 mm², which is tilted by 45° with respect to the horizontal direction (i.e. the dispersion direction of experiments in Bragg-diffraction geometry). Thus it represents the topography, i.e. a map of the local reflectivity of the mirror, which appears very homogeneous. Moreover, the X-ray CCD allows an absolute determination of the detected X-ray signal. Taking into account the known transmission of the Al-filters and quantum efficiency of the CCD, we obtain that the reflected beam contains typically (1.3 - 1.5) x $10^5$ K$_\alpha$-photons/pulse (with pre-pulse and a laser main pulse energy of 110 mJ on target).

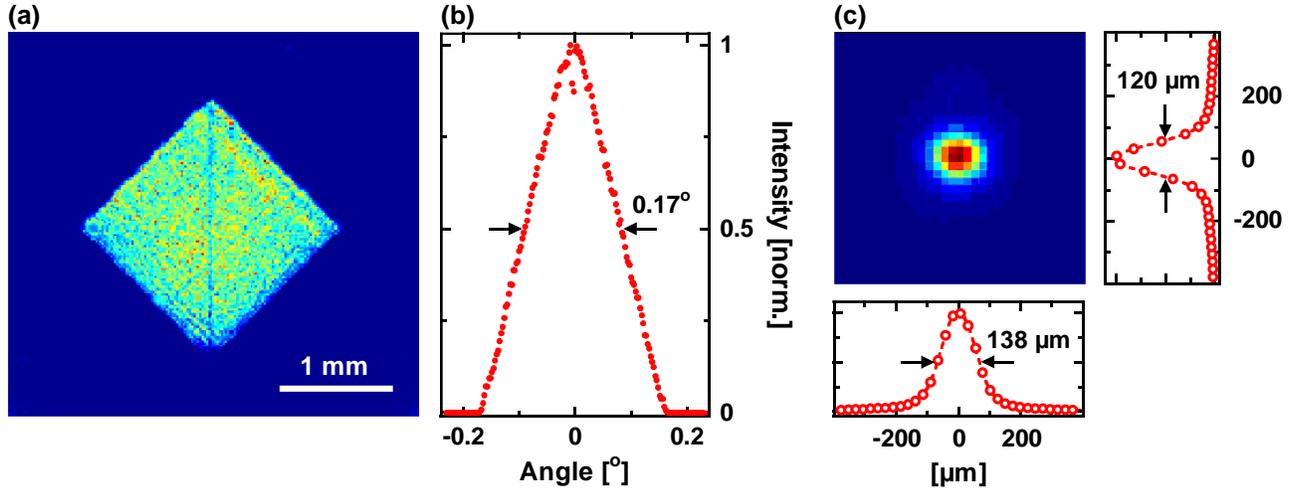

FIG. 4. (a) False color representation of the topography of the multilayer X-ray mirror measured a few cm behind the mirror. (b) Horizontal intensity profile as a function of convergence angle after integration of the image data shown in (a) in vertical direction. (c) Intensity distribution in the focus of the mirror in false color representation (upper left). The graphs at the bottom and to the right represent horizontal and vertical, respectively, cross sections revealing an X-ray spot size (FWHM) of 138 μm (horizontal) and 120 μm (vertical).

The beam exhibits a quadratic shape of 1.6 x 1.6 mm$^2$, which is tilted by 45º with respect to the horizontal direction. Since the distance of the detector to the focus point is known, the measured spatial intensity distribution in Fig. 5(a) allows to calculate the angular properties of the focused beam. The measured size corresponds, therefore, to a small convergence angle of 1.6x10$^{-5}$ sr (0.23º x 0.23º) and thus a quasi-collimated beam. Moreover, in the current setup the horizontal direction represents the dispersion plane for experiments in Bragg-diffraction geometry. The resulting angular intensity profile (after integration along the vertical direction) is shown in Fig. 5(b). As expected, it is triangularly shaped and exhibits a FWHM of 0.17º.

Panel (c) in Fig. 4 depicts measurements of the intensity distribution in the image/focal point of the mirror. It reveals a distortion free, smooth distribution and a slightly elliptical spot with a FWHM of 138 μm horizontally and 120 μm vertically (compare the cross sections in horizontal (bottom) and vertical (right) directions). This agrees with the expectations when considering the mirror magnification of 5 and the 25 μm FWHM of the laser focus on the Cu tape.

**C. Ionization chamber detector**

A low repetition rate system employing high energy laser pulses for X-ray generation has the advantage that the same accumulated diffraction signal can be measured with less X-ray pulses. This comes at the expense that fluctuations of the X-ray flux delivered to the sample are averaged over a low number of events thus leading to a larger measurement error for the diffraction signal.

Therefore, a pulse-to-pulse normalization scheme is desirable. For this purpose, we use a small ionization chamber which is positioned after the X-ray mirror so that only the $K_\alpha$-radiation, which is captured by the mirror and delivered to the sample, is monitored. The ICD is filled with Ar at a pressure of about 1 bar. A fraction of the $K_\alpha$-radiation reflected from the mirror is absorbed by photo-ionizing the neutral Ar atoms. A total transmission of the ICD of 72% is measured, in agreement with the expected transmission value of 1.6 cm Ar at 1 bar and a total thickness of 16 μm of Kapton foil, which is used as cell windows. To collect the charge a high bias voltage is applied (~1900 kV). A high speed current amplifier with a gain of $10^5$ is used to match the output signal of the ICD to the input of a gated charge-integrating analogue-to-digital converter (ADC), which is used to digitize the charge signals.

The generated charge is directly proportional to the incidence X-ray flux as demonstrated by the data depicted in Fig. 5(a). It shows the dependence of the ICD-signal as a function of the transmitted X-ray flux. The latter has been measured with a phosphor-based X-ray area detector (see below) and each data point represents an average over 300 X-ray pulses.

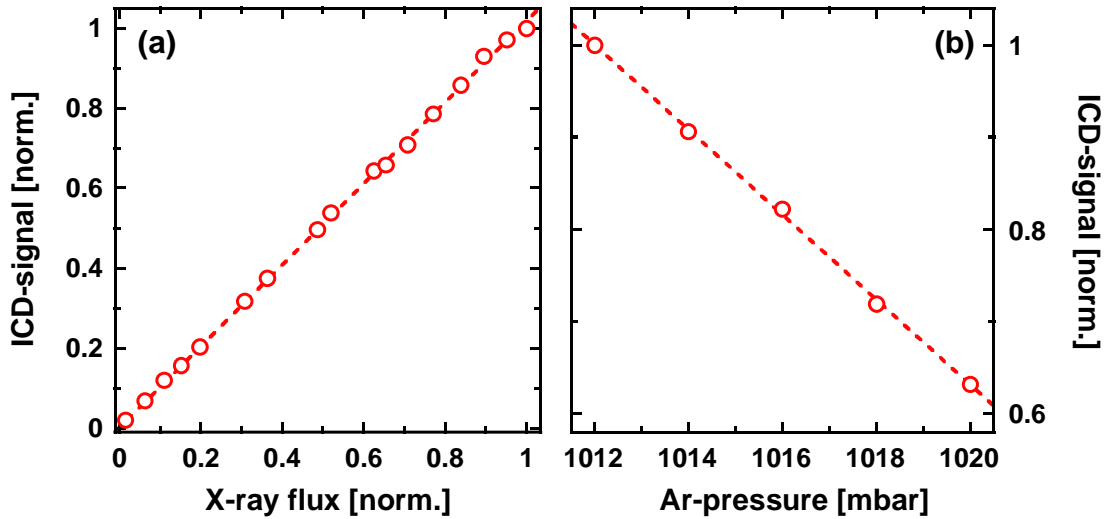

FIG. 5. (a) ICD-signal as a function of the X-ray flux transmitted through the ICD (both quantities have been normalized to the corresponding maximum value). Each data point represents the average over 300 X-ray pulses. (b) ICD-signal (normalized) as a function of Ar-pressure in the cell.

The data in Fig. 5(a) clearly show that the ICD-signal can be used to normalize the diffraction signal. However, it turned out, the generated signal exhibits a strong dependence on the Ar-pressure inside the ICD, as shown in Fig. 6(b). Pressure changes of less than 10 mbar lead to a change of the ICD-signal of almost 40%. Therefore, to keep the pressure inside the ICD constant and also to reduce the gas flow, a pressure control system, consisting of two pinhole valves at the entrance and exit, respectively, and a pressure gauge, has been added. It allows to stabilize the pressure inside the ICD to better than ±0.1 mbar over "short" time-scales of about 30 min. However, over longer time-scales (hours) we observe drifts of the ICD-signal which we attribute to slight changes of the pressure. Measures to mitigate these long-term drift effects will be discussed in the section E.

**D. Sample environment**

The relatively large distance between the (housing of the) X-ray mirror and its image/focus point of 40 cm provides great flexibility with respect to the sample environment. In our setup we have realized two different sample stages dedicated for Bragg-diffraction type measurements under ambient conditions (i.e. in air, no heating or cooling option).

One is a relatively compact sample manipulator, which is used for experiments in a *reversible* excitation regime, where excitation fluences are sufficiently low so that no permanent sample modifications are induced by a single pulse and a pulse-to-pulse sample exchange is, therefore, not necessary.

Fig. 6(a) shows a photo of this sample stage, which has three linear degrees of motion (x, y, z). x (5 cm travel range) and y (3 cm travel range) correspond to motions in the sample plane and thus are used to adjust the position of the X-ray spot on the sample. The z-axis (5cm travel range) allows to bring the sample surface in the focus of the X-ray mirror. In addition, a single-circle goniometer with 100 mm diameter is mounted horizontally underneath the x-y-z stage and used to adjust the angle of incidence ($\theta$) of the X-ray beam, e.g. adjustment of the Bragg-angle, while a smaller rotation stage is mounted on the y-stage to control the azimuthal orientation ($\varphi$) of the sample, which is for example needed to observe asymmetric reflections. All stages are stepper motor driven.

The second sample stage has been constructed for experiments in the *irreversible* excitation regime, i.e. excitation fluences are so high that even illumination with a single optical pump pulse leads to permanent modifications of the sample. As such the sample has to be moved between consecutive exposures over distances of a few hundred μm (multiples of the pump beam spot size). This requires to move larger samples at velocities of a few cm/s with high accuracy.

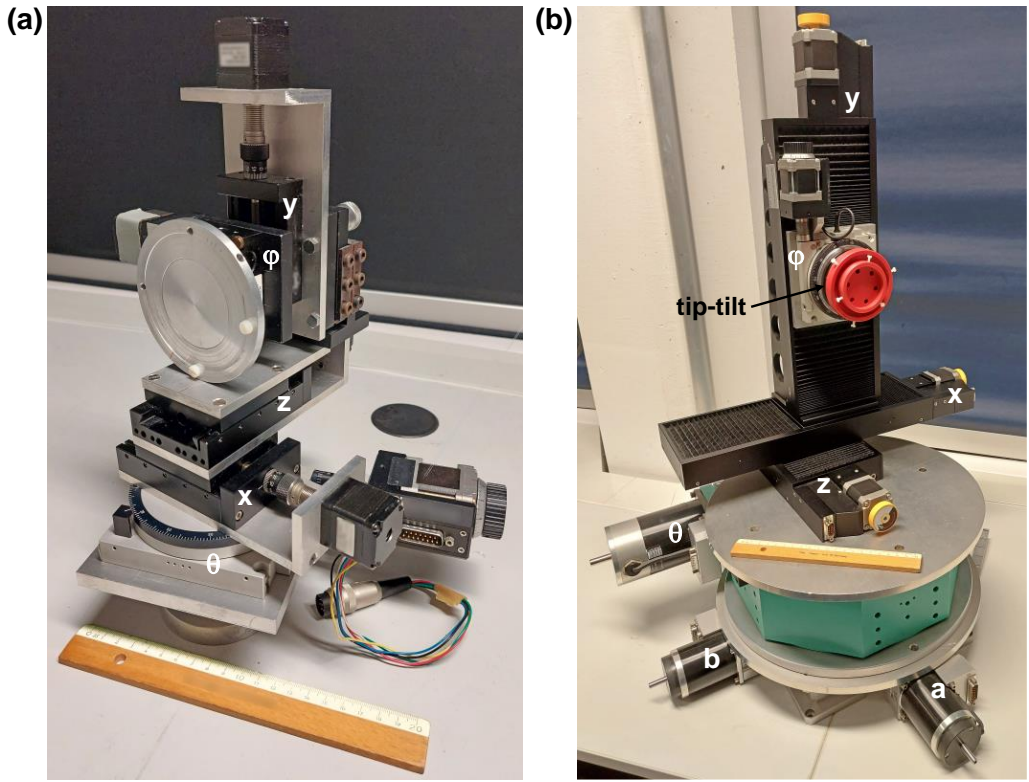

FIG. 6. Two sample stages for diffraction experiments at ambient conditions. (a): "Compact" sample stage designed for experiments under reversible excitation conditions, i.e. excitation fluences that do not lead to permanent modifications of the irradiated sample. (b) "Large" sample stage designed for high-fluence experiments, where laser irradiation leads to permanent changes of the sample and which, therefore, require sample exchange between two consecutive exposures. The different degrees of freedom are discussed in the main text.

Fig. 6(b) shows this sample stage, which is much larger and exhibits more degrees of freedom. A high capacity two axis translation stage with 15 mm motion range for both axes (labelled a and b) represents the base. The motion axes are oriented parallel and perpendicular to the X-ray probe beam direction. On this stage a single-circle goniometer with a diameter of 290 mm is mounted horizontally and used to adjust the angle of incidence ($\theta$) of the X-ray beam. The two-axis stage underneath allows to bring the rotation axis of the goniometer exactly into the X-ray focus. The goniometer carries a three axis sample mover with two axes (x and z) parallel to the goniometer plane, and one axis (y) perpendicular to it. The x- and y-axis (155 mm travel range) correspond to motions in the plane of the sample, and z (50 mm travel range) perpendicular to it. Therefore, z is used to bring the sample surface into the center of rotation, while x and y allow to change the position on the sample. On the y-stage another single-circle goniometer is mounted with the rotation axis parallel to z and thus perpendicular to the sample surface. It allows to adjust the azimuthal orientation ($\varphi$) of the sample. Again all these stages are stepper motor driven. In addition, a manual tip-tilt stage is used to precisely orient the sample surface parallel to the x-y motion plane, which is required to maintain spatial overlap between pump and probe and to keep the sample in the center of rotation for large motion distances in x and y.

Both sample stages can be easily modified for a Debye-Scherrer transmission geometry by removing the φ-stage and replacing it with a sample mount, which is oriented perpendicular to the y-stage.

Other sample environments are possible and, in principle, easy to implement. For example, in-vacuum sample stages combined with heating and/or cooling would allow experiments at variable base temperatures.

**E. Detectors and signal normalization**

Depending on the particular application, three different X-ray detectors are available to record the diffraction signals. The first is a direct-detection, thinned, back-illuminated, slow-scan CCD detector (Princeton Instruments MTE: 1300B). The detector area is 26.8 x 26 mm$^2$ with 1340 x 1300 pixels and 20 x 20 μm$^2$ pixel size. While the quantum efficiency of this detector is rather low (18 %) for Cu K$_\alpha$-radiation it provides single-photon sensitivity, low noise and dark current (thus allowing longer integration times), and 16 bit dynamic range. Moreover, it can be used as an X-ray spectrometer with an energy resolution of about 0.24 keV (see Fig, 3), when operated in photon counting mode.

The second detector is also an area detector, but uses an indirect-detection scheme. The incident X-rays are converted by a scintillator into visible photons, which are then amplified by a MCP and detected by a CCD. The available device (Photonic Science Gemstar HS) has an active area of 32 x 24 mm$^2$ with 1392 x 1040 pixels and 23 x 23 μm$^2$ effective pixel size. The scintillator output is 1:1 fiber-coupled to a gated, 40 mm diameter MCP-based image intensifier with variable gain. Coupling between the intensifier output screen and the CCD is achieved by a tapered fiber-optic, so that the CCD is viewing a rectangular area with a diagonal of 40mm on the intensifier input. This detector has a high quantum efficiency of 85 % for Cu K$_\alpha$-radiation and the variable MCP gain allows adaptation to the X-ray input level, reaching almost single photon sensitivity at high gain. Moreover, with 2x2 binning 10 Hz operation and thus single-pulse detection is possible.

The third detector is a single-channel detector based on a large-area (10 mm diameter) Si avalanche photodiode (API 394-70-74-591). It has a high sensitivity down to the single-photon level, much faster read-out time allowing pulse-to-pulse data acquisition even at repetition rates much higher than 10 Hz, compact size and can be operated at room temperature (no cooling). Using the direct-detection CCD for calibration a quantum efficiency of the APD for Cu K$_\alpha$ radiation of approx. 40 % was measured, in principal agreement with previously published data[45,46]. Since the APD-signals are recorded and digitized with the same charge-integrating ADC as used for the ICD (and the laser reference diode), the APD output signal is first fed into a pulse-shaping amplifier (charge amplification of 11).

To characterize the performance of the APD signals have been recorded for low X-ray input levels (1 – 3 photons/pulse) over a few thousand pulses for different bias voltages. Some results are depicted in Fig. 7. Fig. 7(a) shows a histogram of the number of measured events as a function of the APD-signal (given in ADC-units - adu) for a bias-voltage of 1.9 kV. The signal value distribution exhibits a multi-peak structure as it is expected for a Si-detector with its rather well defined relation between the charge generated by a single X-ray photon and the photon energy (i.e. one Cu $K_\alpha$ photon produces $3.66 \times 10^{-4}$ pCb). While the first peak corresponds to the dark current background (no photon detected, proven by blocking the incident X-rays), the second, third, fourth, … peak correspond to one, two, three, … detected $K_\alpha$-photons.

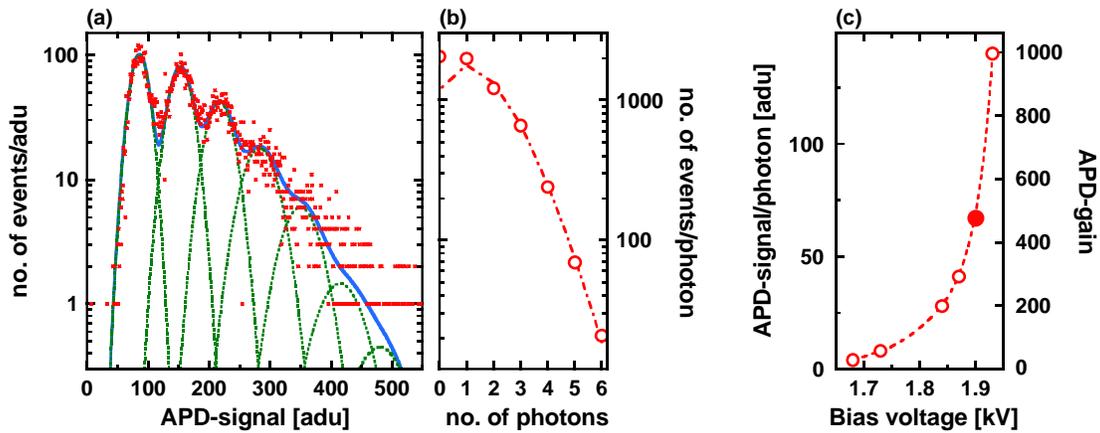

FIG. 7. Histogram of the APD-signal (given in ADC-units) at low incident X-ray flux measured over approx. 10 min. for a bias voltage of 1.9 kV. Red squares: measured data; blues solid curve: fit by a superposition of Gaussian functions (green dashed curves). The first, second, third, … peak correspond to zero (i.e. dark current), one, two, … detected $K_\alpha$-photons, respectively. (b) Integrated number of events as a function of $K_\alpha$-photon number. Red open circles: experimental data; red dashed-dotted curve: fit by a Poisson distribution with an average number of 1.5 $K_\alpha$-photons per pulse. (c) APD-signal (in ADC-units) per $K_\alpha$-photon as a function of bias voltage (left axis) and APD-gain (right axis). The red dashed curve represents a guide-to-the-eye); the solid red dot corresponds to a bias voltage of 1.9 kV, for which data are presented in (a) and (b).

A multi-peak fit (Gaussian) was applied to separate the individual contributions and to determine in particular the position of each peak. The latter exhibits a linear dependence as a function of photon number resulting in an average signal of 67 adu per $K_\alpha$-photon. Integrating the number of events under each peak the total number of events as a function of photon number can be obtained, as depicted in Fig. 7(b). The measured data (open circles) can be described by a Poisson-distribution with on average 1.5 $K_\alpha$-photons detected per pulse (red dashed-dotted curve).

Fig. 7(c) shows on the left ordinate the APD-signal per photon (in adu) as a function of bias voltage (the red solid dot corresponds to a bias voltage of 1.9 kV, for which data are presented in Fig. 8(a) and (b)). Knowing the number of primary electrons produced in the APD per $K_\alpha$-photon, the gain of the pulse-shaping amplifier, and the sensitivity of the ADC (1 adu per 0.313 pCb) the APD-signal per $K_\alpha$-photon can be converted into the APD-gain as shown on the right ordinate. In experiments with weakly diffraction samples resulting in typical signals of 5 - 30 detected photons per pulse we operate the APD with bias voltages of 1.9 kV and above, corresponding to gain values of approx. 500 – 1000.

As already pointed out above, at low repetition rate and measurements with a comparably low total number of pump-probe events, normalization is key to achieve a high measurement accuracy. For higher repetition (1 kHz) different schemes have been discussed in the literature[17,44,47,48], which depend also on the type of detector (e.g. its read-out speed). In particular, for single-channel detectors, like the APD discussed here, normalization is usually achieved by chopping the optical pump beam at half the system repetition rate. As such the detected signal is a sequence of alternating probe-only and pump-probe events and the accumulated signal of all pump-probe events is normalized to the accumulated signal of all probe-only events. Holtz et al. have analyzed in detail the noise-characteristics using this *chopping*-scheme for normalization and demonstrated almost shot-noise-limited performance[47].

For our low-repetition rate source we have compared the *chopping*-scheme with the direct normalization using the ICD as a pulse-to-pulse reference. For this purpose, we recorded the APD-signal for different signal levels from a few detected photons per pulse up to approx. 100 detected photons per pulse for about an hour (30000 – 40000 pulses), with and without the ICD in the beam path. Due to the finite X-ray transmission of the ICD (72%) the signals in the measurements without the ICD in the beam path were a factor of $1/0.72 \approx 1.4$ higher than with the ICD.

These datasets have then been divided into $N$ bins of different number of events $M$ (corresponding to a total acquisition time $t_a = M/10\ Hz$ for one data point for example in a pump-probe experiment). Using the *chopping*-scheme, in each bin the APD-signal of every second event is normalized to the APD-signal of the previous event. Then all the single-pulse ratios in a bin are averaged and their normalized standard deviation $\Delta S/S$ is determined. Using direct normalization, the APD-signal of each pulse in a bin is normalized to the corresponding ICD-signal, then all APD/ICD-ratios are averaged over all events in a bin and subsequently $\Delta S/S$ is determined.

Some results are depicted in Fig. 8, which shows the average normalized standard deviation $\langle \Delta S/S \rangle_N$ for different signal levels $n_{ph}$ (as given in the insets of Fig. 8) as a function of $t_a$; (a): *chopping*-scheme; (b) direct normalization with ICD. The error bars represent the standard deviation of $\Delta S/S$, and the different colors encode equivalent data sets with and without the ICD in the beam-path; e.g. red: 7 detected photons per pulse without ICD in (a) correspond to 5 detected photons per pulse with ICD in (b). The dashed-dotted curves represent the expected $\Delta S/S$ assuming that the accuracy is limited by counting statistics and thus the total number of detected photons $n_{tot}$ per data point: $\Delta S/S = 1/\sqrt{n_{tot}}$.

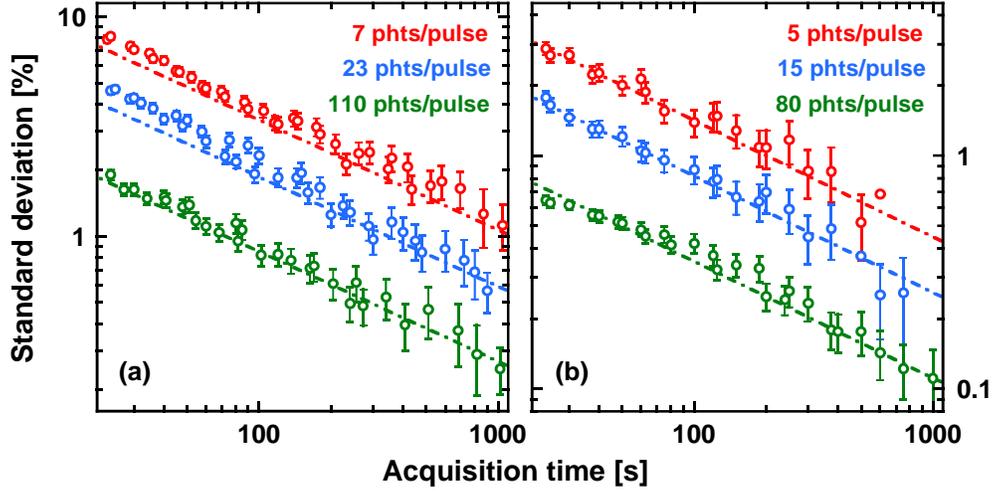

FIG. 8. Standard deviation of the measured diffraction signal for different signal levels (average number of detected photons per pulse) as a function of the acquisition time. (a): Results for the *chopping*-scheme (see text). (b) Results when using direct normalization with the ICD as reference detector. The dashed-dotted curves represent the expected standard deviation when the measurement accuracy is limited by counting statistics (i.e. total number of detected photons).

First of all, the data in Fig. 8 clearly show, that from the probe-side our setup allows measurements with an accuracy only limited by counting statistics. Moreover, despite the 40% higher single-pulse signal $n_{ph}$ for the chopping-scheme (no ICD in the beam path), a much larger acquisition time of almost a factor of 6 is needed to achieve the same accuracy as in a corresponding measurement with direct normalization (e.g. 1% at 23 phts/pulse with *chopping*: ≈360 s; 1% at 15 phts/pulse with direct normalization: ≈65 s).

This result is again fully in line with the expectations of a counting-statistics limited measurement. For the *chopping*-scheme, both, the actual pump-probe signal and the probe-only normalization signal, are limited by counting statistics and both are accumulated over only 50% of the total acquisition time $t_a^{ch}$. Therefore, with $n_a^{ch}$ the corresponding average number of detected photons per pulse $\Delta S/S$ can be expressed as:

$$\Delta S/S = \frac{2}{\sqrt{10\,Hz \cdot (t_a^{ch}/2) \cdot n_a^{ch}}} \qquad (1)$$

For direct normalization with the ICD as reference it must be noted that the ICD-signal corresponds to almost $3 \times 10^4$ $K_\alpha$-photons per pulse and, therefore, its counting statistics does not limit the accuracy. Moreover, for an equivalent experimental situation the average number of detected photons $n_a^d$ with the ICD in place is $n_a^d = 0.72 \cdot n_a^{ch}$. This yields:

$$\Delta S/S = \frac{1}{\sqrt{10\ Hz \cdot t_a^d \cdot 0.72 \cdot n_a^{ch}}} \qquad (2)$$

The ratio $t_a^{ch}/t_a^d$ to achieve the same $\Delta S/S$ with the two normalization schemes is:

$$t_a^{ch}/t_a^d = \frac{4 \cdot 10\ Hz \cdot t_a^d \cdot 0.72 \cdot n_a^{ch}}{10\ Hz \cdot (t_a^{ch}/2) \cdot n_a^{ch}} = 8 \cdot 0.72 = 5.76 \qquad (3)$$

Therefore, direct normalization with the ICD provides clear advantages over the chopping scheme, namely either a higher accuracy at a given acquisition time or a much shorter acquisition time for the same accuracy.

## III. APPLICATION EXAMPLES

Laser-induced structural changes lead to distinct changes of diffraction signals, namely of the shape and position of Bragg peaks due to strain, and of the integrated diffraction efficiency due to variations of the structure factor. The scientific applications, for which the presented setup can be used best, are determined by the properties of the X-ray probe beam delivered to the sample by the multilayer X-ray mirror: (i) moderate monochromaticity of $2 \times 10^{-3}$ (full $K_\alpha$-bandwidth), and (ii) a relatively small convergence angle of 0.23° (quasi-collimated). For experiments on single-crystalline samples in Bragg-geometry the setup lacks angular resolution and does not allow to observe small angular shifts or a subtle broadening of Bragg-peaks, making it unsuitable for the investigation of strain-related phenomena. In contrast, the high X-ray flux delivered to the sample and the demonstrated counting-statistics limited accuracy enable precise investigations of processes that lead only to structure factor changes and thus to variations of the integrated diffraction efficiency. Moreover, the quasi-collimated beam allows for Debye-Scherrer-type diffraction experiments in transmission geometry. Specific examples for both applications will be presented in the following.

### A. Detection of coherent optical phonons in single-crystalline, thin Bismuth films

A specific case where the structural changes upon laser-irradation lead to changes of the structure factor alone is the excitation of coherent optical phonons in Bismuth. It is well known that in Bi the fully symmetric $A_{1g}$ optical mode can be excited in a coherent fashion through a mechanism called *Displacive Excitation of Coherent Phonons* (DECP)[49]. The $A_{1g}$ phonon modulates the distance of Bi atoms along the body-diagonal (111-direction) of the rhombohedral unit cell and thus the structure factor of all Bragg-reflections with momentum transfer along (111). Time resolved X-ray diffraction has been used frequently to address this phenomenon[11,50-53,22].

The high per-pulse X-ray flux of the current setup, which enables high accuracy measurements with a rather low number of pump-probe events and thus low accumulated optical dose (see II), allowed us to study DECP in Bismuth in a high pump fluence regime, which had not been accessible before. While a detailed discussion of the results will be the topic of a separate publication, Fig. 10, depicts as an example data obtained on a 50 nm thick, (111)-oriented, epitaxial Bi-film grown on a (111)-oriented Si substrate[54].

Fig. 9(a) shows the angular dependence ($\theta$-$2\theta$-scan) of the diffraction signal. It exhibits the (111) peaks of the Bi-film and the Si-substrate, respectively (normalized to the peak signal of Bi-peak). The Si-peak is by a factor of 20 stronger than the Bi-reflection which is expected because it is a bulk material and not a thin film. The width of the Si- and Bi-peaks are $\approx 0.18°$ (FWHM) and $\approx 0.3°$ (FWHM), respectively. The measured width is determined by the angular width of the X-ray beam (0.17°), its spectral width ($\approx 2.5 \times 10^{-3}$), and the natural width of the particular Bragg-reflection. The spectral width makes only a small contribution ($\Delta\theta \approx 0.03° - 0.04°$). Since the Si-substrate exhibits very high crystalline quality its natural width is very small and can be neglected. Correspondingly, the measured width is essentially equal to the angular width of the X-ray probe beam at the sample. In contrast, the measured Bi-peak is significantly broader than both, the natural width of a 50 nm Bi-film ($\approx 0.08°$), and the angular width of the X-ray beam. We attribute the additional broadening to the mosaic structure of the Bi-film. In the time-resolved diffraction experiment the angle was adjusted to the maximum of the Bi-curve and the APD-detector positioned accordingly.

Fig. 9(b) shows results of a time-resolved measurement with an incident pump fluence of 3 mJ/cm$^2$. The diffraction signal, recorded on a single-pulse basis by the APD, corresponds to 10 – 20 detected photons per pulse and was integrated over 2 min per delay point to ensure that, in principle, an accuracy of approximately 1 % could be reached (compare Fig. 9(b)). To account for any drifts on longer time scales (e.g. induced by small variations of the pressure in the ICD or in the spatial overlap of optical pump and X-ray probe on the sample) we adapted a "fast" scanning technique introduced for all-optical pump-probe experiments at high repetition rates[55]. Instead of accumulating the signal for a given pump-probe delay time over 2 minutes at once and varying $\Delta t$ sequentially, the full delay range was scanned multiple times (here 12x) with a short acquisition time per delay point (here 10 seconds).

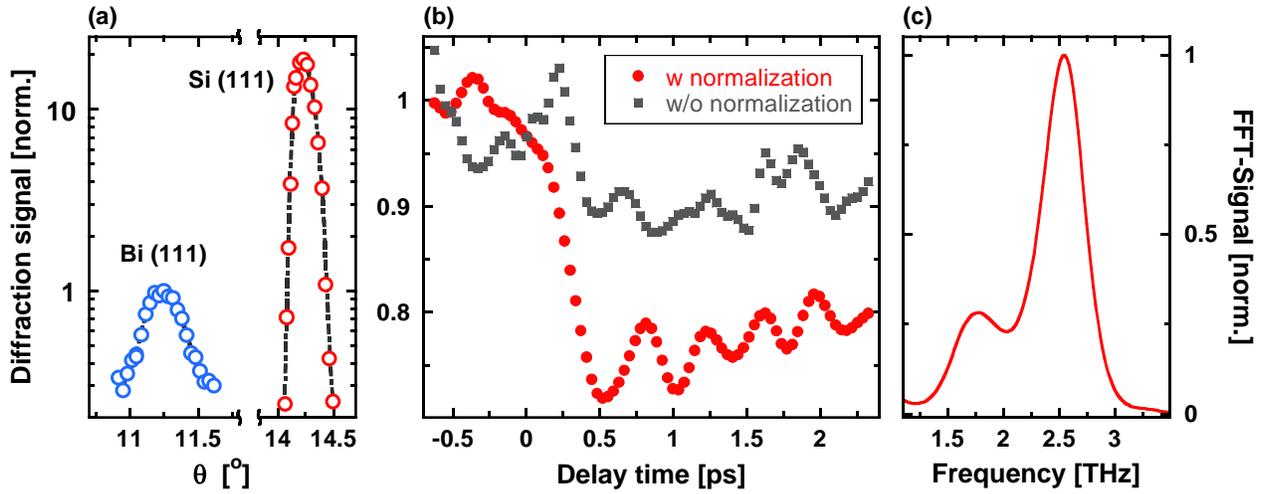

FIG. 9. (a) Normalized θ−2θ-scan (rocking curve) of the 50 nm Bi (111) film grown on a Si (111) substrate (left Bi; right Si). (b) Diffraction signal of the Bi (111)-reflection as a function of time delay Δt between laser pump and X-ray probe normalized to the (averaged) diffraction signal measured at negative Δt for an incident pump fluence of 3 mJ/cm$^2$; red dots: with normalization to the ICD-signal; grey squares: without normalization. (c) FFT (squared amplitude) of the measured signal for Δt > 0.3 ps.

The grey squares represent the obtained result without normalization of the measured diffraction signal to the corresponding ICD-signal. The red dots are obtained with normalization. For each scan the diffraction signal was normalized on a pulse-to-pulse basis to the ICD-signal, averaged, and then re-normalized to the average signal measured for all negative Δt (i.e. representing the diffraction efficiency of the unexcited sample). Finally, all the renormalized delay scans were averaged.

This comparison clearly demonstrates the need for and the effectiveness of the described normalization procedure. Despite a drop of the average diffraction signal at positive delay times no clear transient changes are visible without normalization, in particular no oscillatory behavior which may be attributed to the generation of coherent phonons. In contrast, the normalized data (red data points) exhibit a much larger drop of the average diffraction signal at positive time delays and, in particular, clear signal oscillations caused by the coherent excitation of the $A_{1g}$ optical mode.

Fig. 9(c) depicts the squared amplitude of the Fourier-transform of the oscillatory part of the signal measured at positive time delays, revealing a dominant peak at a frequency of 2.53 THz. This value is significantly lower than the $A_{1g}$ phonon frequency of unexcited Bi of 2.92 THz at room temperature[56]. Similar red-shifts have been observed in many time-resolved optical[57-59] and X-ray diffraction[11,50,51,53,22] experiments on laser-excited Bi and can be readily explained by the changes of the inter-atomic potential (shift of the potential minimum and softening) upon optical/electronic excitation, which represent the basis of the DECP mechanism.

One may be tempted to attribute the second peak at 1.8 THz to the similarly red-shifted $E_g$ optical phonon modes of Bi. While these modes correspond to atomic motion perpendicular to the (111)-direction and the measured diffraction signal is not directly sensitive to this, it has been suggested that anharmonic interactions between the $A_{1g}$ and the $E_g$ modes can lead to additional atomic motion along the (111) direction with the frequency of the $E_g$-mode[57,60]. However, we refrain from interpreting the 1.8 THz peak in this way, since we observe signal changes even at negative time delays, which we have to attribute to residual signal fluctuations/drifts that cannot be normalized by our procedure.

**B. Debye-Scherrer diffraction at poly-crystalline metal foils**

One of the key properties of the X-ray beam delivered to the sample is its low convergence angle. This quasi-collimated X-ray probe beam in combination with the relatively low spectral bandwidth makes this setup also suited for Debye-Scherrer-type diffraction experiments on poly-crystalline materials and powders in transmission geometry using an area detector[16,17,61,62]. Although the scattering signal is not localized in an intense diffraction spot, but distributed over a ring, this scheme has a number of advantages: (i) greater sample flexibility - no single-crystalline/epitaxial samples are required; (ii) simplicity - no precise sample adjustments (i.e. Bragg-angle) are necessary; (iii) several Bragg-peaks can be recorded simultaneously.

Since we recently demonstrated the applicability of our setup for time-resolved Debye-Scherrer diffraction by analyzing the picosecond acoustic response of a 200 nm Au-film upon fs laser excitation[62], we will discuss here only briefly some results of static diffraction experiments.

Fig. 10 shows in the top row detector images measured on a 20 µm thick Cu-foil (a) and a free-standing 200 nm Au film supported by a Ni-mesh (b). Because of its larger size we used the phosphor-based, MCP-amplified area detector (see section II.E) at high gain. The detector was placed close to the sample (normal distance $l_0 = 35\ mm$) and at an oblique angle ($\alpha = 28°$) to the direction of the X-ray beam, to record as many diffraction orders as possible (covered range of scattering angles: 35° - 85°). In both cases the signal was accumulated over 3000 pulses (5 min exposure time).

The bottom row shows the diffraction signal $I(q)$ as a function of momentum transfer $q = \frac{4\pi}{\lambda} \cdot \sin\left(\frac{\theta}{2}\right)$ ($\theta$: scattering angle) after azimuthal integration of the diffraction images. The individual Bragg-reflections are labelled with their respective Miller-indices (the scattering pattern of the Au-films contains weak scattering contributions from the Ni mesh).

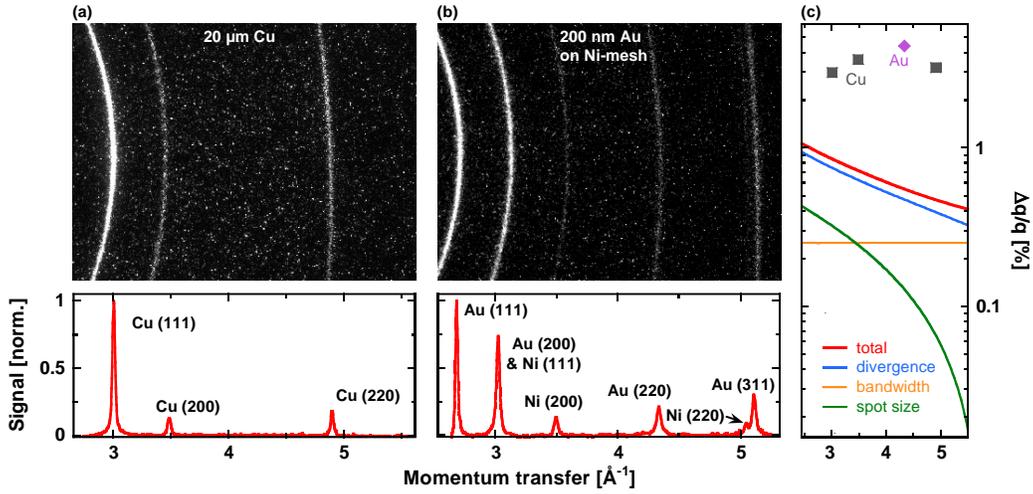

FIG. 10. Diffraction data obtained on poly-crystalline materials in transmission geometry. Top row: Detector images from (a) a 20 μm thick Cu-foil, and (b) a free-standing, 200 nm thick Au-film supported by a Ni-mesh (in both cases accumulated over 3000 pulses). Bottom row: Corresponding scattering profiles $I(q)$ as a function of momentum transfer $q = \frac{4\pi}{\lambda} \cdot \sin\left(\frac{\theta}{2}\right)$ ($\theta$: scattering angle) after azimuthal integration of the diffraction images. (c) Different contributions to the total normalized momentum resolution $\Delta q/q$ (red curve) due to (i) divergence of the X-ray beam (blue curve), (ii) the X-ray spot size (green curve), and (iii) the bandwidth (orange curve). The grey squares and the violet diamond represent the measured (relative) peak widths of the three measured Cu reflections and the (220) Au reflection, respectively.

An important performance measure is the momentum resolution $\Delta q/q$, which is determined in our setup by three different contributions: (i) the divergence of the beam $\Delta\theta_{div} = 0.17°$, (ii) the projected X-ray spot size on the sample, and (iii) the spectral bandwidth. While (iii) makes a fixed contribution of $(\Delta q/q)_{bw} = \Delta E/E = 2.5 \times 10^{-3}$, (i) depends on the scattering angle $\theta$, and (ii) on the scattering angle $\theta$ and the sample detector distance $l_0$.

By differentiating Bragg's law contribution (i) can be determined as $(\Delta q/q)_{div} = \Delta\theta_{div}/[2 \cdot \tan(\theta/2)]$. With $\theta_0 = 90° - \alpha$ contribution (ii) can be written as $(\Delta q/q)_{foc} = \frac{d \cdot \cos\theta \cdot \cos(\theta - \theta_0)}{2 \cdot l_0 \cdot \tan(\theta/2)}$. The total momentum resolution $\Delta q/q$, is then given by the convolution of (i) to (iii):

$$\frac{\Delta q}{q} = \sqrt{\left(\frac{\Delta q}{q}\right)^2_{div} + \left(\frac{\Delta q}{q}\right)^2_{foc} + \left(\frac{\Delta q}{q}\right)^2_{bw}} \qquad (4)$$

Quantitative estimates for our setup and the given detector geometry ($l_0$ and $\alpha$) are depicted in Fig. 10(c). The blue, green, and orange curves show the contributions due to the beam divergence $(\Delta q/q)_{div}$, the spectral bandwidth $(\Delta q/q)_{bw}$, and the X-ray spot size $(\Delta q/q)_{foc}$, respectively, while the red curve represents the total momentum resolution $\Delta q/q$. These results clearly show that over the covered *q*-range the momentum resolution is essentially limited by the beam divergence, while the other effects make only minor contributions.

Fig. 10(c) also shows the measured relative peak widths of the three Cu reflections (grey squares) and of the (220) Au reflections (violet diamond; the (200)- and the (311)-reflection could not be properly analyzed due to the nearby Ni-reflections, as the (111)-reflection, which is partially cut), which are significantly larger than the limits set by the setup. We attribute this to the finite size of the crystalline grains. Using Scherrer's equation (with form-factor K = 1) we estimate grain sizes of about 20 – 25 nm in line with the observation of smooth diffraction rings.

## IV. SUMMARY

In summary, we have presented a compact and modular, and thus very flexible, setup for time-resolved X-ray diffraction using ultrashort Cu $K_\alpha$ X-ray pulses from a laser-produced plasma. The X-ray source employs a thin Cu-tape as target, which is excited by high energy (>100 mJ) laser pulses with a wavelength of 800 nm and a pulse duration of 110 fs at a repetition rate of 10 Hz. Using a pre-pulse scheme, an X-ray flux of about $10^{10}$ $K_\alpha$ photons per pulse (into the full solid angle) was achieved of which about $10^5$ are transferred onto the investigated sample in a quasi-collimated beam with a beam diameter of about 130 µm (FWHM) by means of a multilayer Montel mirror. An ionization chamber detector placed behind the X-ray mirror provides a normalization signal and allows to achieve counting-statistics-limited measurement accuracy. Different area and single detectors as well as different sample stages can be used to adapt the setup to specific requirements. The characteristics and the overall performance of the setup have been demonstrated with test experiments in Bragg- as well in Debye-Scherrer geometry.

## ACKNOWLEDGMENTS

Financial support by the Deutsche Forschungsgemeinschaft (DFG, German Research Foundation) through the Collaborative Research Centre 1242 (project number 278162697) is gratefully acknowledged. We thank M. Kammler for preparation of the Bi sample.

## AVAILABILITY OF DATA

The data that support the findings of this study are available within the article.

**REFERENCES**

[1] R. Schoenlein, T. Elsaesser, K. Holldack, Z. Huang, H. Kapteyn, M. Murnane, and M. Woerner, Philos. Trans. R. Soc. A **377**, 20180384 (2019)).

[2] T. Elsaesser and M. Woerner, J. Chem. Phys. **140**, 020901 (2014).

[3] C. Bostedt, S. Boutet, D. M. Fritz, Z. Huang, H. J. Lee, H. T. Lemke, A. Robert, W. F. Schlotter, J. J. Turner, and G. J. Williams, Rev. Mod. Phys. **88**, 015007 (2016).

[4] M. Dunne, Nat. Rev. Mater. **3**, 290 (2018).

[5] H. N. Chapman, Annu. Rev. Biochem. **88**, 35 (2019).

[6] C. Rischel, A. Rousse, I. Uschmann, P.-A. Albouy, J.-P. Geindre, P. Audebert, J.-C. Gauthier, E. Fröster, J.-L. Martin, and A. Antonetti, Nature **390**, 490 (1997).

[7] C. Rose-Petruck, R. Jimenez, T. Guo, A. Cavalleri, C. W. Siders, F. Rksi, J. A. Squier, B. C. Walker, K. R. Wilson, and C. P. J. Barty, Nature **398**, 310 (1999).

[8] C. W. Siders, A. Cavalleri, K. Sokolowski-Tinten, C. Toth, T. Guo, M. Kammler, M. H. V. Hoegen, K. R. Wilson, D. V. D. Linde, and C. P. J. Barty, Science **286**, 1340 (1999).

[9] A. Rousse, C. Rischel, S. Fourmaux, I. Uschmann, S. Sebban, G. Grillon, P. Balcou, E. Förster, J. P. Geindre, P. Audebert, J. C. Gauthier, and D. Hulin, Nature **410**, 65 (2001).

[10] K. Sokolowski-Tinten, C. Blome, C. Dietrich, A. Tarasevitch, M. Horn von Hoegen, D. von der Linde, A. Cavalleri, J. Squier, and M. Kammler, Phys. Rev. Lett. **87**, 225701 (2001).

[11] K. Sokolowski-Tinten, C. Blome, J. Blums, A. Cavalleri, C. Dietrich, A.Tarasevitch, I. Uschmann, E. Förster, M. Kammler, M. Horn-von Hoegen, and D. von der Linde, Nature **422**, 287 (2003).

[12] M. Bargheer, N. Zhavoronkov, Y. Gritsai, J. C. Woo, D. S. Kim, M. Woerner, and T. Elsaesser, Science **306**, 1771 (2004).

[13] C. v Korff Schmising, M. Bargheer, M. Kiel, N. Zhavoronkov, M. Woerner, T. Elsaesser, I. Vrejoiu, D. Hesse, and M. Alexe, Phys. Rev. Lett. **98**, 257601 (2007).

[14] F. Quirin, M. Vattilana, U. Shymanovich, A.-E. El-Kamhawy, A. Tarasevitch, J. Hohlfeld, D. von der Linde, and K. Sokolowski-Tinten, Phys. Rev. B **85**, 020103 (2012).

[15] J. Stingl, F. Zamponi, B. Freyer, M. Woerner, T. Elsaesser, and A. Borgschulte, Phys. Rev. Lett. **109**, 147402 (2012).

[16] V. Juve, M. Holtz, F. Zamponi, M. Woerner, T. Elsaesser, and A. Borgschulte, Phys. Rev. Lett. **111**, 217401 (2013).

[17] M. Holtz, C. Hauf, A. H. Salvador, R. Costard, M. Woerner, and T. Elsaesser, Phys. Rev. B **94**, 104302 (2016)

[18] J. Pudell, A. A. Maznev, M. Herzog, M. Kronseder, C. H. Back, G. Malinowski, A. von Reppert, and M. Bargheer, Nat. Commun. **9**, 3335 (2018).

[19] A. von Reppert, L. Willig, J.-E. Pudell, S. P. Zeuschner, G. Sellge, F. Ganss, O. Hellwig, J. A. Arregi, V. Uhlíř, A. Crut, and M. Bargheer, Science Adv. **6**, eaba1142 (2020).

[20] M. Mattern, A. von Reppert, S. Zeuschner, J.-E. Pudell, F. Kühne, D. Diesing, M. Herzog, and M. Bargheer, Appl. Phys. Lett. **120**, 092401 (2022).

[21] S. Priyadarshi, I. González-Vallejo, C. Hauf, K. Reimann, M. Woerner, and T. Elsaesser, Phy. Rev. Lett. **128**, 136402 (2022).

[22] A. Koç, I. González-Vallejo, M. Runge, A. Ghalgaoui, K. Reimann, L. Kremeyer, F. Thiemann, M. Horn von Hoegen, K. Sokolowski-Tinten, M. Woerner, and T. Elsaesser, Phys. Rev. B **107**, L180303 (2023).

[23] N. Zhavoronkov, Y. Gritsai, M. Bargheer, M. Woerner, T. Elsaesser, F. Zamponi, I. Uschmann, and E. Förster, Opt. Lett. **30**, 1737 (2005).

[24] F. Zamponi, Z. Ansari, C. v. Korff Schmising, P. Rothhardt, N. Zhavoronkov, M. Woerner, T. Elsaesser, M. Bargheer, T. Trobitzsch-Ryll, M. Haschke, Appl Phys A **96**, 51 (2009).

[25] J. Weisshaupt, V. Juvé, M. Holtz, S. Ku, M. Woerner, T. Elsaesser, S. Ališauskas, A. Pugžlys, and A. Baltuška, Nat. Photonics **8**, 927 (2014).

[26] A. Koç, C. Hauf, M. Woerner, L. von Grafenstein, D. Ueberschaer, M. Bock, U. Griebner, and T. Elsaesser, Opt. Lett. **46**, 210 (2021).

[27] S. Fourmaux and J. C. Kieffer, Appl. Phys. B **122**, 162 (2016).

[28] Y. Azamoum, R. Clady, A. Ferré, M. Gambari, O. Utéza, and M. Sentis, Opt. Lett. **43**, 3574 (2018).

[29] P. Gibbon, E. Förster, Plasma Phys. Control. Fusion **38**, 769 (1996).

[30] C. Reich, P. Gibbon, I. Uschmann and E. Förster, Phys. Rev. Lett. **84**, 4846 (2000).

[31] M. Bargheer, N. Zhavoronkov, R. Bruch, H. Legall, H. Stiel, M. Woerner, and T. Elsaesser, Appl. Phys. B **80**, 715 (2005).

[32] U. Shymanovich, M. Nicoul, K. Sokolowski-Tinten, A. Tarasevitch, C. Michaelsen, and D. von der Linde, Appl. Phys. B **92**, 493 (2008).

[33] U. Shymanovich, M. Nicoul, W. Lu, S. Kähle, A. Tarasevitch, K. Sokolowski-Tinten, and D. von der Linde, Rev. Sci. Instrum. **80**, 083102 (2009).

[34] F. Zamponi, Z. Ansari, M. Woerner, and T. Elsaesser, Opt. Express **18**, 947 (2010).

[35] R. Rathore, V. Arora, H. Singhal, T. Mandal, J. A. Chakera, and P. A. Naik, Laser Part. Beams **35**, 442 (2017).

[36] M. Schollmeier, T. Ao, E. S. Field, B. R. Galloway, P. Kalita, M. W. Kimmel, D. V. Morgan, P. K. Rambo, J. Schwarz, J. E. Shores, I. C. Smith, C. S. Speas, J. F. Benage, and J. L. Porter, Rev. Sci. Instrum. **89**, 10F102 (2018).

[37] M. Afshari, P. Krumey, D. Menn, M. Nicoul, F. Brinks, A. Tarasevitch, and K. Sokolowski-Tinten, Struct. Dyn. **7**, 014301 (2020)

[38] W. Lu, M. Nicoul, U. Shymanovich, A. Tarasevitch, P. Zhou, K. Sokolowski-Tinten, D. von der Linde, M. Mašek, P. Gibbon, and U. Teubner, Phys. Rev. E **80**, 026404 (2009).

[39] M. Montel, Optica Acta **1**, 117 (1954).

[40] A. Lübcke, Ph.D. thesis, Faculty of Physics and Astronomy, Friedrich-Schiller-Universität Jena, Jena, 2007.

[41] I. Gilath, D. Salzmann, M. Givon, M. Dariel, L. Kornblit, T. Bar-Noy, J. Mat. Sc. **23**, 1825 (1988).

[42] L. Labate, M. Galimberti, A. Giulietti, D. Giulietti, L.A. Gizzi, P. Tomassin, G. Di Cocco, Nucl. Instr. Meth. A **495**, 148 (2002).

[43] F. Zamponi, T. Kämpfer, A. Morak, I. Uschmann, and E. Förster, Rev. Sci. Instrum. **76**, 116101 (2005).

[44] D. Schick, A. Bojahr, M. Herzog, C. von Korff Schmising, R. Shayduk, W. Leitenberger, P. Gaal, and M. Bargheer, Rev. Sci. Instrum. **83**, 025104 (2012).

[45] L. M. P. Fernandes, F. D. Amaro, A. Antognini, J. M. R. Cardoso, C. A. N. Conde, O. Huot, P. E. Knowles, F. Kottmann, J. A. M. Lopes, L. Ludhova, C. M. B. Monteiro, F. Mulhauser, R. Pohl, J. M. F. dos Santos, L. A. Schaller, D. Taqqud and J. F. C. A. Velosoe, J. of Instrumentation (JINST) **2**, 08005 (2007).

[46] L. M. P. Fernandes, J. A. M. Lopes, J. M. F. dos Santos, P. E. Knowles, L. Ludhova, F. Mulhauser, F. Kottmann, R. Pohl, D. Taqqu, IEEE Trans. Nucl. Sc. **51**, 1575 (2004).

[47] M. Holtz, C. Hauf, J. Weisshaupt, A. H. Salvador, M. Woerner, and T. Elsaesser, Struct. Dyn. **4**, 054304 (2017).

[48] B. Freyer, J. Stingl, F. Zamponi, M.Woerner, and T. Elsaesser, Opt. Express **19**, 15506 (2011).

[49] H. J. Zeiger, J. Vidal, T. K. Cheng, E. P. Ippen, G. Dresselhaus, and M. S. Dresselhaus, Phys. Rev. B 45, 768 (1992).

[50] D. M. Fritz, D. A. Reis, B. Adams, R. A. Akre, J. Arthur, C. Blome, P. H. Bucksbaum, A. L. Cavalieri, S. Engemann, S. Fahy, R. W. Falcone, P. H. Fuoss, K. J. Gaffney, M. J. George, J. Hajdu, M. P. Hertlein, P. B. Hillyard, M. Horn-von Hoegen, M. Kammler, J. Kaspar, R. Kienberger, P. Krejcik, S. H. Lee, A. M. Lindenberg, B. McFarland, D. Meyer, T. Montagne, E. D. Murray, A. J. Nelson, M. Nicoul, R. Pahl, J. Rudati, H. Schlarb, D. P. Siddons, K. Sokolowski-Tinten, Th. Tschentscher, D. von der Linde, and J. B. Hastings, Science **315**, 633 (2007).

[51] S. L. Johnson, P. Beaud, C. J. Milne, F. S. Krasniqi, E. S. Zijlstra, M. E. Garcia, M. Kaiser, D. Grolimund, R. Abela, and G. Ingold, Phys. Rev. Lett. **100**, 155501 (2008).

[52] S. L. Johnson, P. Beaud, E. Vorobeva, C. J. Milne, É. D. Murray, S. Fahy, and G. Ingold, Phys. Rev. Lett. 102, 175503 (2009).

[53] S. W. Teitelbaum, T. C. Henighan, H. Liu, M. P. Jiang, D. Zhu, M. Chollet, T. Sato, É. D. Murray, S. Fahy, S. O'Mahony, T. P. Bailey, C. Uher, M. Trigo, and D. A. Reis, Phys. Rev. B **103**, L180101 (2021).

[54] M. Kammler, M. Horn-von Hoegen, Surf. Sci. **576**, 56 (2005).

[55] G. C. Cho, W. Kütt, and H. Kurz, Phys. Rev. Lett. **65**, 764 (1990).

[56] J. S. Lannin, J. M. Calleja, and M. Cardona, Phys. Rev. B **12**, 585 (1975).

[57] M. Hase, M. Kitajima, S. Nakashima, and K. Mizoguchi, Phys. Rev. Lett. **88**, 067401 (2002).

[58] É. D. Murray, D. M. Fritz, J. K. Wahlstrand, S. Fahy, and D. A. Reis, Phys. Rev. B **72**, 060301(R), (2005).

[59] S. W. Teitelbaum, T. Shin, J. W. Wolfson, Y.-H. Cheng, I. J. Porter, M. Kandyla, and K. A. Nelson, Phys. Rev. X **8**, 031081 (2018).

[60] E.S. Zijlstra, L. L. Tatarinova, and M. E. Garcia, Phys. Rev. B **74**, 220301(R) (2006).

[61] U. Shymanovich, M. Nicoul, W. Lu, S. Kähle, A. Tarasevitch, K. Sokolowski-Tinten, and D. von der Linde, Rev. Sci. Instrum. **80**, 083102 (2009).

[62] F. Zamponi, Z. Ansari, M. Woerner, and T. Elsaesser, Opt. Express **18**, 947 (2010).

[63] W. Lu, M. Nicoul, U. Shymanovich, F. Brinks, M. Afshari, A. Tarasevitch, D. von der Linde, and K. Sokolowski-Tinten, AIP Advances **10**, 035015 (2020).